\documentclass{article}

\PassOptionsToPackage{numbers, compress}{natbib}

\usepackage[preprint]{neurips_2024}

\usepackage[utf8]{inputenc} 
\usepackage[T1]{fontenc}    
\usepackage{hyperref}       
\usepackage{url}            
\usepackage{booktabs}       
\usepackage{amsfonts}       
\usepackage{nicefrac}       
\usepackage{microtype}      
\usepackage{xcolor}         
\usepackage{float}
\usepackage{graphicx}
\title{Agentic LLM Workflows for Generating Patient-Friendly Medical Reports}

\author{
Malavikha Sudarshan$^{1}$, Sophie Shih$^{2}$, Estella Yee$^{2}$, Alina Yang$^{2}$, John Zou$^{3}$, \\ 
\textbf{Cathy Chen$^{4}$, Quan Zhou$^{5}$, Leon Chen$^{5}$, Chinmay Singhal$^{5}$ and George Shih$^{6}$} \\
$^{1}$Department of Electrical Engineering and Computer Sciences, University of California, Berkeley, CA, USA \and $^{2}$Stuyvesant High School, NY, USA \and
$^{3}$Department of Computer Science, Brown University, RI, USA \and
$^{4}$Stern School of Business, New York University, NY, USA \\
$^{5}$MD.ai, NY, USA \and
$^{6}$Department of Radiology, Weill Cornell Medicine, NY, USA \\
\parbox{\textwidth}{\centering
\texttt{malavikhasudarshan@berkeley.edu, \{sshih60, eyee60, ayang6\}@stuy.edu, john\_zou@brown.edu, hc2845@nyu.edu, \{quan, leon, chinmay\}@md.ai, george@cornellradiology.org} 
} 
}

\begin{document}

\maketitle

\begin{abstract}
The application of Large Language Models (LLMs) in healthcare is expanding rapidly, with one potential use case being the translation of formal medical reports into patient-legible equivalents. Currently, LLM outputs often need to be edited and evaluated by a human to ensure both factual accuracy and comprehensibility, and this is true for the above use case. We aim to minimize this step by proposing an agentic workflow with the Reflexion framework, which uses iterative self-reflection to correct outputs from an LLM. This pipeline was tested and compared to zero-shot prompting on 16 randomized radiology reports. In our multi-agent approach, reports had an accuracy rate of 94.94\% when looking at verification of ICD-10 codes, compared to zero-shot prompted reports, which had an accuracy rate of 68.23\%. Additionally, 81.25\% of the final reflected reports required no corrections for accuracy or readability, while only 25\% of zero-shot prompted reports met these criteria without needing modifications. These results indicate that our approach presents a feasible method for communicating clinical findings to patients in a quick, efficient and coherent manner whilst also retaining medical accuracy. The codebase is available for viewing at \url{http://github.com/malavikhasudarshan/Multi-Agent-Patient-Letter-Generation}.
\end{abstract}


\section{Background}

The 21st Century Cures Act grants patients the right to access their electronic health record data, and since its implementation, the number of patients accessing their Electronic Health Records (EHRs) before the ordering provider has increased significantly \citep{pollock2024}. While intended to improve transparency and promote a shared flow of information, this increased level of accessibility can often lead to patient anxiety, misinterpretation and confusion when reading jargon-filled medical reports that they are not the primary audience for \citep{gerber2022}. Radiology reports are a prime example of these; mostly intended for referring physicians, when abnormal or ambiguous results are received by patients before discussion with their physician, the impact can often be more harmful than beneficial \citep{winget2016}. To address this, the creation of patient-friendly letters that simplify complex medical information has been explored \citep{smolle2021}. These letters aim to explain medical terms clearly, ensure factual accuracy, and also maintain a compassionate and reassuring tone. 

In recent years, Large Language Models (LLMs) have been increasingly leveraged in healthcare applications, from producing discharge summaries \citep{zaretsky2024} to structured radiology reports \citep{liu2023}. In applying generative artificial intelligence to the creation of patient-friendly reports, the pipeline can be made more efficient and patients can have access to more meaningful and legible letters \cite{doo2023, park2024}. Most current developments invoke zero-shot prompting to create a patient-friendly version of a medical report included in the input prompt, where the LLM’s internal representations are relied upon to produce a suitable letter, and no template or example output is provided in the prompt for guiding the structure, style or comprehensiveness of the generated letters \citep{cork2024, roberts2023}. Through this method, LLMs often generate outputs that need to be manually reviewed or go through alternative mechanisms to be critiqued and improved before being delivered to the patient. One research study concluded that 80.4\% (\texttt{n = 41}) of tested patient-friendly LLM-generated summaries of medical reports required editing before being released to patients \citep{berigan2024}. 

Our goal was to develop an agentic pipeline where verification would be minimized, and where patient letters would be evaluated for both accuracy and readability before being released.

\section{Methods}

Agentic workflows are iterative and consist of several intermediary steps performed in addition to LLM prompting, as opposed to non-agentic or zero-shot/few-shot prompts which consist of a single input and a single output \citep{ng2024}. The former approach means that multiple agents can be leveraged, and they are often structured similar to professional businesses, where each agent plays a specific role in the organization. Addition-by-Subtraction collaboration is one example of a multi-agent method, where one agent provides information and the other removes unnecessary details and provides feedback \citep{wu2024}.

Agentic workflows allow for reinforcement learning through reflection \citep{ng2024}, and can utilize chain-of-thought prompting by appending reflected feedback at the end of the next prompt. We leveraged an existing framework, Reflexion \citep{shinn2023}, which incorporates verbal reinforcement into its iterative refinement process. Typically, agents receive feedback from their environment in a simple form, like a binary signal (e.g., success/failure) or a scalar value (a numerical score). Reflexion agents take this basic feedback and translate it into a more detailed, verbal form—a textual summary that explains the feedback in natural language. This verbal feedback is then added to the context of the following prompt, and acts as a 'semantic' gradient signal, meaning that it provides the agent with specific, meaningful directions on how to improve. 

Our implementation prompts an LLM to generate a specific number of patient-friendly letters based on a formal medical report. The accuracy and readability of each generated letter is calculated and weighted appropriately, and the Reflexion model is then used to run a certain number of self-reflection trials and output the letter that it considers to be optimal at the end of this. Reflexion has three separate legs – AlfWorld (for decision-making problems), HotPotQA (for reasoning on single-step iterations) and Programming (for programming tasks using interpreters and compilers). We used AlfWorld, as decision-making made the most sense when prompting for multiple letters and asking for the most optimal output.

The original medical report can either be provided as an argument, or, as we presented at the Society for Imaging Informatics in Medicine (SIIM) 2024 Hackathon, be pulled from an EHR server. Our integration involved extracting one of the five medical reports available on the SIIM Fast Healthcare Interoperability Resources (FHIR) server and pushing our results back onto the server. Manually including the medical report in the input was also later tested on 15 other test radiology reports of various modalities: Computed Tomography (CT), Magnetic Resonance Imaging (MR), and Ultrasound (US) (see Figs. 1, 2 and 3 in the Appendix). These reports differed in length, ranging from 84 to 264 words, and covered a range of medical findings and body parts, including the abdomen, pelvis, chest, head, and lumbar spine.

\begin{figure}
  \centering
  \includegraphics[width=0.68\linewidth]{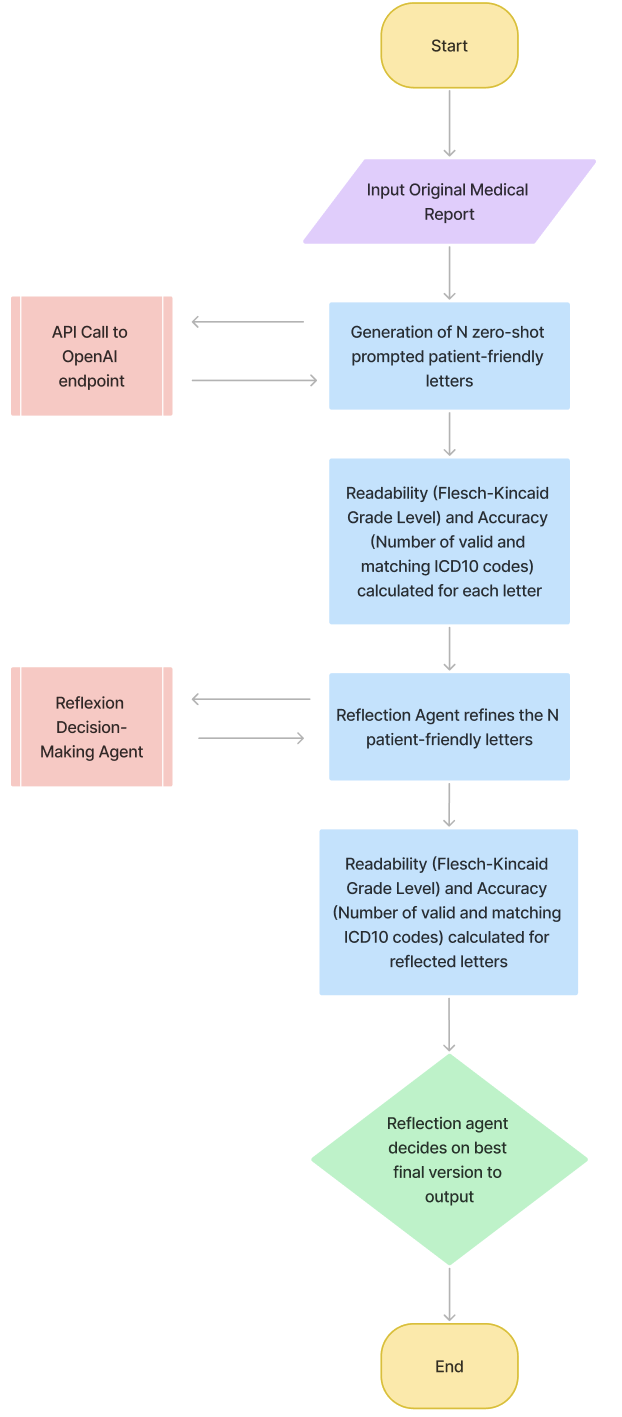}
  \caption{Flowchart of the Multi-Agent Algorithm}
  \label{fig:flowchart}
\end{figure}

Our pipeline (Fig.~\ref{fig:flowchart}) operates as follows: we first make one LLM call to extract the International Classification of Diseases, Tenth Revision (ICD-10) codes from the original report. The temperature is kept at 0 to minimize variance, and these codes are stored to be compared later. A second LLM call is then used to generate a number of patient-friendly reports (\texttt{n=5}, for example) based on the original, and this time we ask the agent to produce ICD-10 codes based on the content of each patient-friendly letter. These ICD-10 codes are verified against the master ICD-10 code database (using the \texttt{simple-icd-10} package \citep{travasci2024}) and the description for each code is also retrieved and compared against the LLM’s output to see if they match. The accuracy of each letter is calculated as the number of validated and identical ICD-10 codes between the patient-friendly version and the original medical report, divided by the total number of ICD-10 codes on the original report. This value should be maximized. Readability is quantified using the Flesch-Kincaid Grade Level. This value is calculated using a predefined formula incorporating the average sentence length, number of syllables and number of words per sentence \citep{kincaid1975} and can be accessed by importing the readability module \citep{readability2024}. The average American’s reading ability is equivalent to a US 8th-grade level \citep{amin2023}. A previous study examining 97,052 radiology reports revealed that only 4\% were at a reading level equal to or below the reading ability of a US 8th-grader \citep{martin2019}, suggesting that much of this information may be unintelligible to a significant portion of the population. Our 16 test reports had an average Flesch-Kincaid Grade Level of 11.03, corresponding to an 11th-grader’s expected level of vocabulary.

A Flesch-Kincaid Grade Level of 6.0 (corresponding to a US 6th-grader’s reading ability) is the recommended level of readability advised by the American Medical Association \citep{weiss2003} and the National Institute of Health \citep{badarudeen2010} for patient-facing medical materials, to allow for greater comprehensibility and accessibility \citep{davis1990}. Each generated patient letter’s overall score is calculated by weighting the readability and accuracy - we wanted to prioritize medical accuracy so opted to compute the score as follows:

\texttt{overall\_score = (readability * 0.3) + (accuracy * 0.7)}

The readability value is standardized to be as close to 6.0 as possible, therefore, we can aim for an \texttt{overall\_score} that has a maximum value of 1.0. Reflexion’s Alfworld module is then used to reflect on the overall\_score, looking to improve both the accuracy and readability of each letter on each iteration. The algorithm outputs the best version of the letter, which is then directly pushed to the linked EHR server for patient access, demonstrating end-to-end integration. The LLM used in our tests was OpenAI’s GPT-4o (gpt-4o-2024-05-13).

\newpage

\section{Results}

Our reflection agent increased the medical accuracy of reports, ensuring that ICD-10 codes were retained in the final patient letter which the zero-shot output sometimes missed. When given an identical medical report, system prompt and user prompt, the reflected output consistently scored higher in terms of accuracy and readability, as well as in the \texttt{overall\_score} measure. Zero-shot prompts were sometimes not professional enough, and even when specified in the prompt that the reading level should match that of a US 6th grader’s, the language used was too juvenile. However, when the reflected agent was used, the final outputs seemed to be consistently more concise, structured and formal.

\begin{figure}[H]
  \centering
  \includegraphics[width=\linewidth]{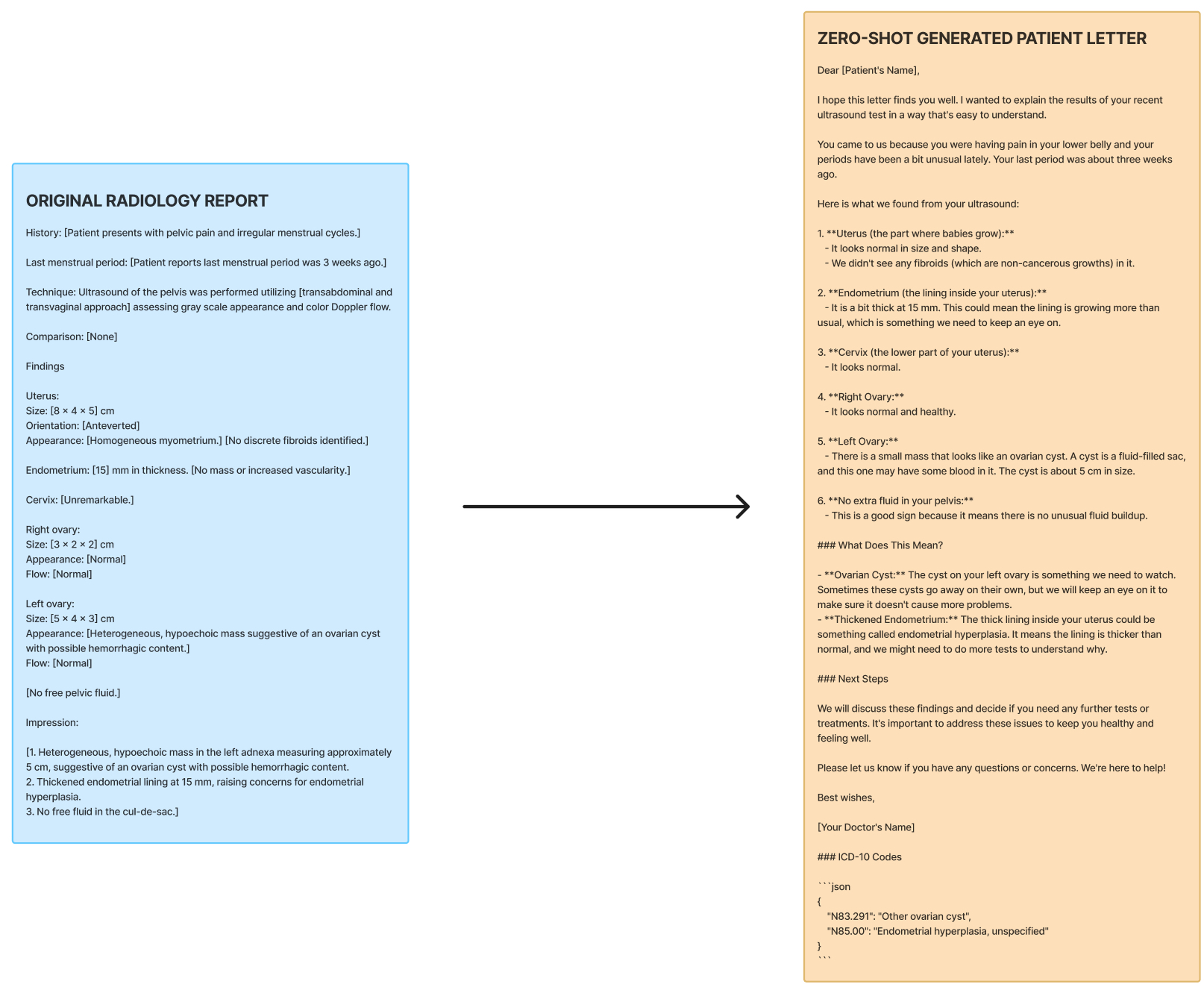}
  \caption{Zero-Shot Generated Patient-Friendly Letter}
  \label{fig:zero-shot}
\end{figure}

\begin{figure}[H]
  \centering
  \includegraphics[width=\linewidth]{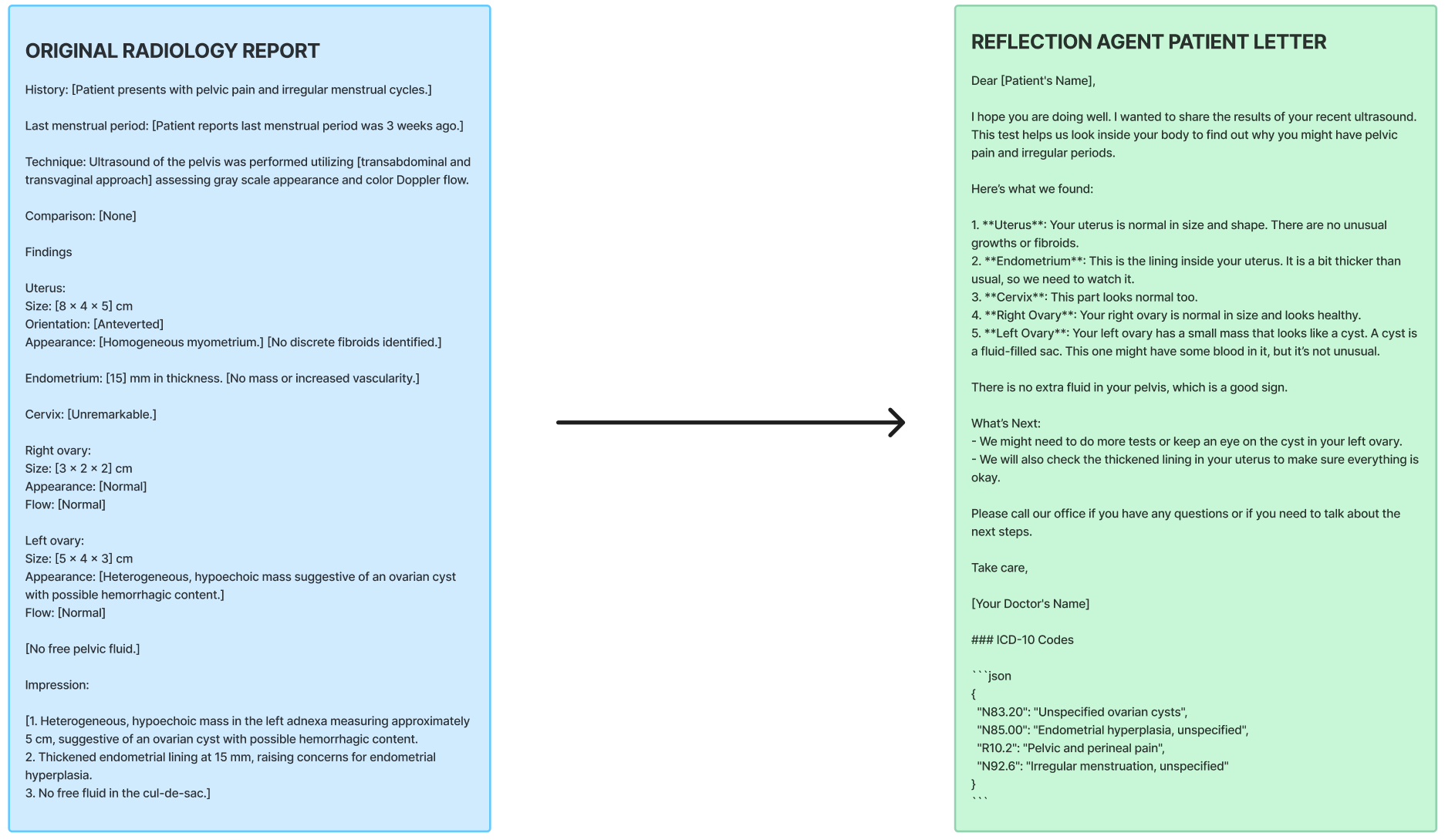}
  \caption{Reflection/Multi-Agent Generated Patient-Friendly Letter}
  \label{fig:reflection}
\end{figure}

In the example below (Figs.~\ref{fig:zero-shot} and ~\ref{fig:reflection}), the same medical report was used to compare a zero-shot (Fig.~\ref{fig:zero-shot}) and reflection agent output (Fig.~\ref{fig:reflection}). With zero-shot prompting, only half of the desired ICD-10 codes were registered, whereas the reflection agent successfully generated all 4 ICD-10 codes. Additionally, the ICD-10 codes generated by the reflection agent precisely matched those from the original medical report, while the codes from the zero-shot report did not.

\begin{figure}[H]
  \centering
  \includegraphics[width=\linewidth]{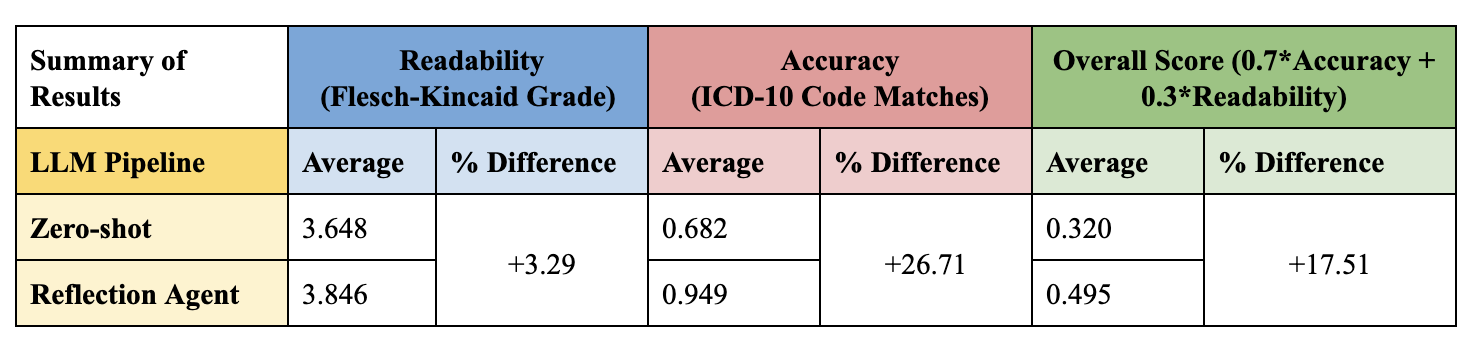}
  \caption{Summary Table of Results}
  \label{fig:summarytable}
\end{figure}

From 16 test radiology reports, zero-shot prompting (using the same original prompt as given in our multi-agent workflow) led to 11/16 patient-friendly versions needing to be edited, whilst our agentic workflow resulted in only 3/16 reports that required modification. We considered ‘modification’ to be any changes in medical factuality (including ICD-10 codes), grammar, punctuation, tonality and readability. On average, accuracy was 26.71\% better, and readability scored 3.29\% higher in the reflected patient letters, compared to the zero-shot letters (Fig.~\ref{fig:summarytable}). This resulted in a 17.51\% increase in \texttt{overall\_score} in reflected letters vs zero-shot generated letters.

\section{Discussion}

The use of ChatGPT and similar LLMs for the generation of patient friendly letters is something that many others in the healthcare space have been experimenting with \citep{zaretsky2024}\citep{roberts2023}\citep{ali2023}\citep{guo2024}\citep{doshi2024}. However, LLMs are known to hallucinate and are extremely sensitive to input, which can often lead to errors in the outputted patient-friendly letter. Additionally, the more complex the medical report, the higher the tendency for LLMs to hallucinate \citep{xu2024}. The inclusion of multiple agents and programmatic prompts\footnote{LLM prompts have been programmed and cannot be altered by users—handles issue of sensitivity to input.} aim to manage the complexity of medical reports, whilst simultaneously minimizing hallucinations. This workflow reduces the need for proofreading, as the patient letters are evaluated for both accuracy and readability before being outputted. 

As part of our accuracy metric, we make use of the \texttt{get\_description(icd10\_code)} \citep{travasci2024} function to verify whether the ICD-10 code definitions match industry-standard for the original and patient-friendly reports. However, as this function uses string matching, it is possible that we may miss out on synonymous definitions or phrases with a few variances in words. A better alternative may be to use fuzzy matching algorithms such as calculating the Levenshtein distance \citep{lee2019}, or looking at the K-Nearest Neighbors \citep{luschow1970} of the two description strings to categorize them, instead of comparing two strings for an exact match. 

One assumption we make is the accuracy of the ICD-10 codes generated by the LLM model (GPT-4o).  In separate human validation tests, we have seen high accuracy and consistency when generating these codes from test radiology reports, so in this study we assume that these generated ICD-10 codes can be trusted.

This is still a very early prototype and can be improved upon in several ways. In the future, we hope to be more inclusive of different reading levels, languages, and medical fields. Currently, we have standardized the level of readability to a 6th grade level; however, it would be beneficial to have a variety of literacy levels available depending on the patient. Additionally, adding the functionality for accurate translation in various languages would significantly enhance communication abilities as well as global applicability and reach. Finally, we are aiming to be applicable to various medical fields outside radiology. 

As of now, our weighting system is based upon readability and accuracy. However, we understand the importance of maintaining a certain level of compassion within these letters. One possible approach is utilizing the PERMA model \citep{butler2016} as a metric for factoring in compassion into our weighting system. The PERMA scale can help our model determine whether a patient letter has the appropriate tone and level of sensitivity. Other additional metrics we are looking into to further enhance patient letters include CDE codes \citep{nih2024}, which can help to accurately convey a patient's treatment process and future action required. 

\section{Conclusion}

Our objective was to find a method of generating patient-friendly radiology reports that would really reduce the need for a medical professional to review and verify them. Although our approach does significantly improve the quality of patient-legible letters, it does not have an 100\% success rate, and therefore cannot eradicate the need for verification completely. However, by significantly reducing the percentage of LLM-generated reports requiring edits—from 68.75\% to 18.75\%—through the incorporation of a multi-agent workflow, we can show that the time spent making changes in medical accuracy and readability will also be substantially decreased. Our method not only enhances the efficiency of report generation but also contributes to the overall goal of making healthcare information more accessible and understandable to patients. This development has strong potential to streamline clinical workflows, lessen the burden on medical professionals and administrators, and improve the patient experience by swiftly providing clearer and more accurate medical information.

\section{Acknowledgements}

The authors would like to thank the SIIM community, most notably Teri M. Sippel Schmidt, Alex Barrington, Tom O’Sullivan, Mohannad Hussain, and those who took the time to provide feedback, for supporting our work.

\newpage

\bibliography{references}


\newpage

\appendix

\section{Appendix}

The following examples display LLM outputs for MR, CT, and Ultrasound test reports, offering straightforward visual comparisons between zero-shot and multi-agent generated patient-friendly reports.

\begin{figure}[H]
  \centering
  \includegraphics[width=\linewidth]{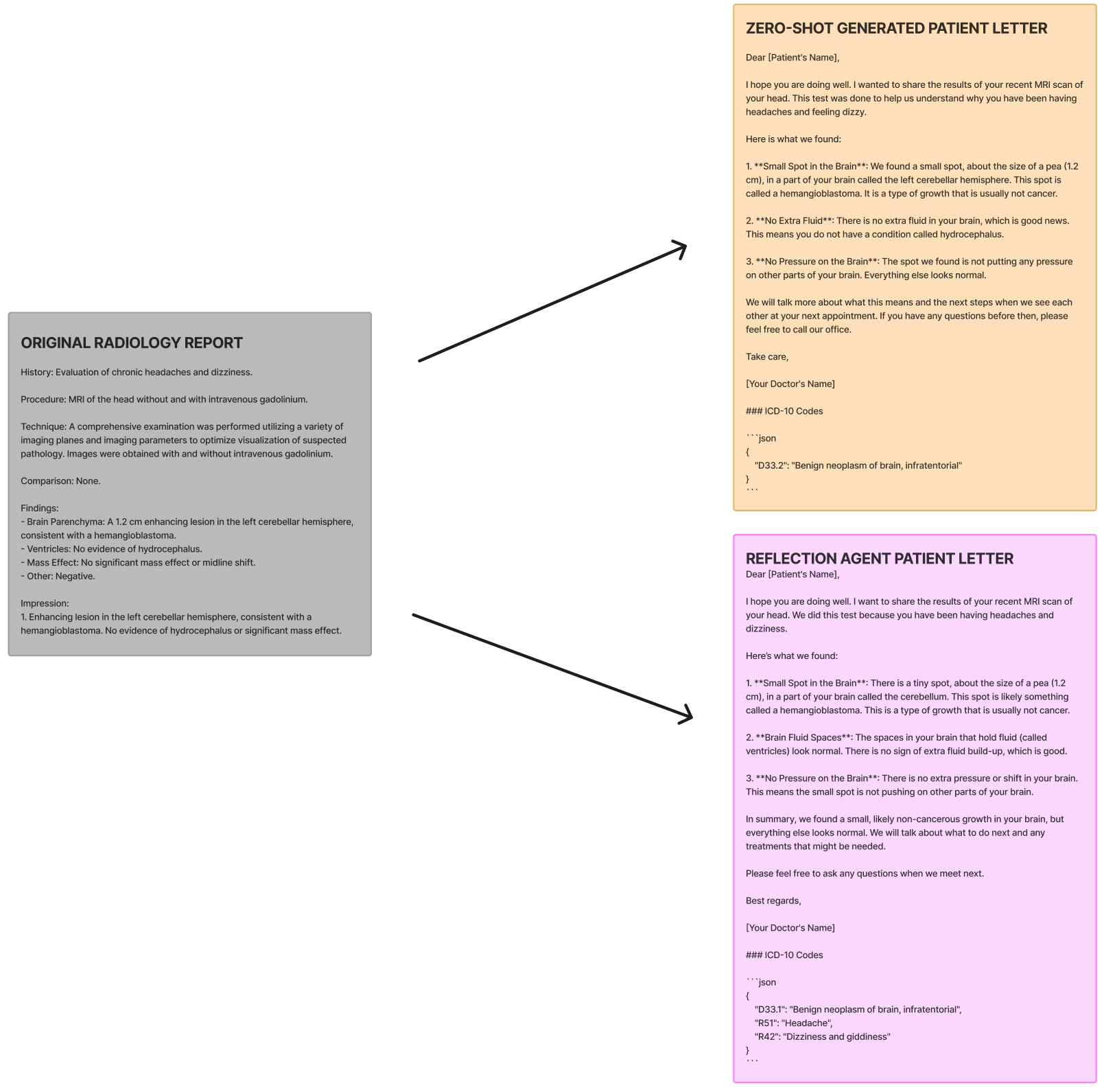}
  \caption{Patient-Friendly Letters generated from an MR Head Report}
  \label{fig:mr_head}
\end{figure}

\begin{figure}
  \centering
  \includegraphics[width=\linewidth]{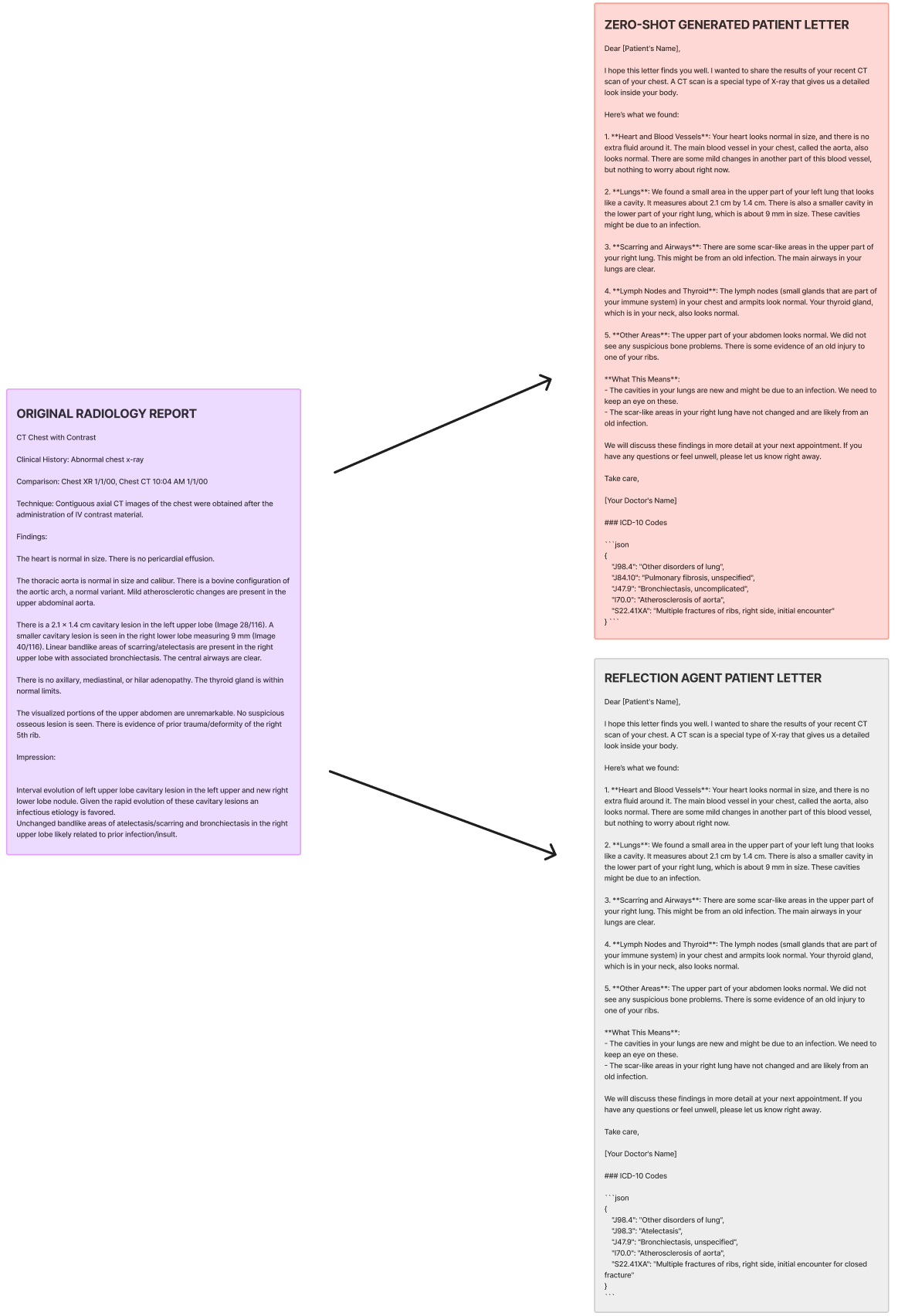}
  \caption{Patient-Friendly Letters generated from a CT Chest Report}
  \label{fig:ct_chest}
\end{figure}

\begin{figure}
  \centering
  \includegraphics[width=\linewidth]{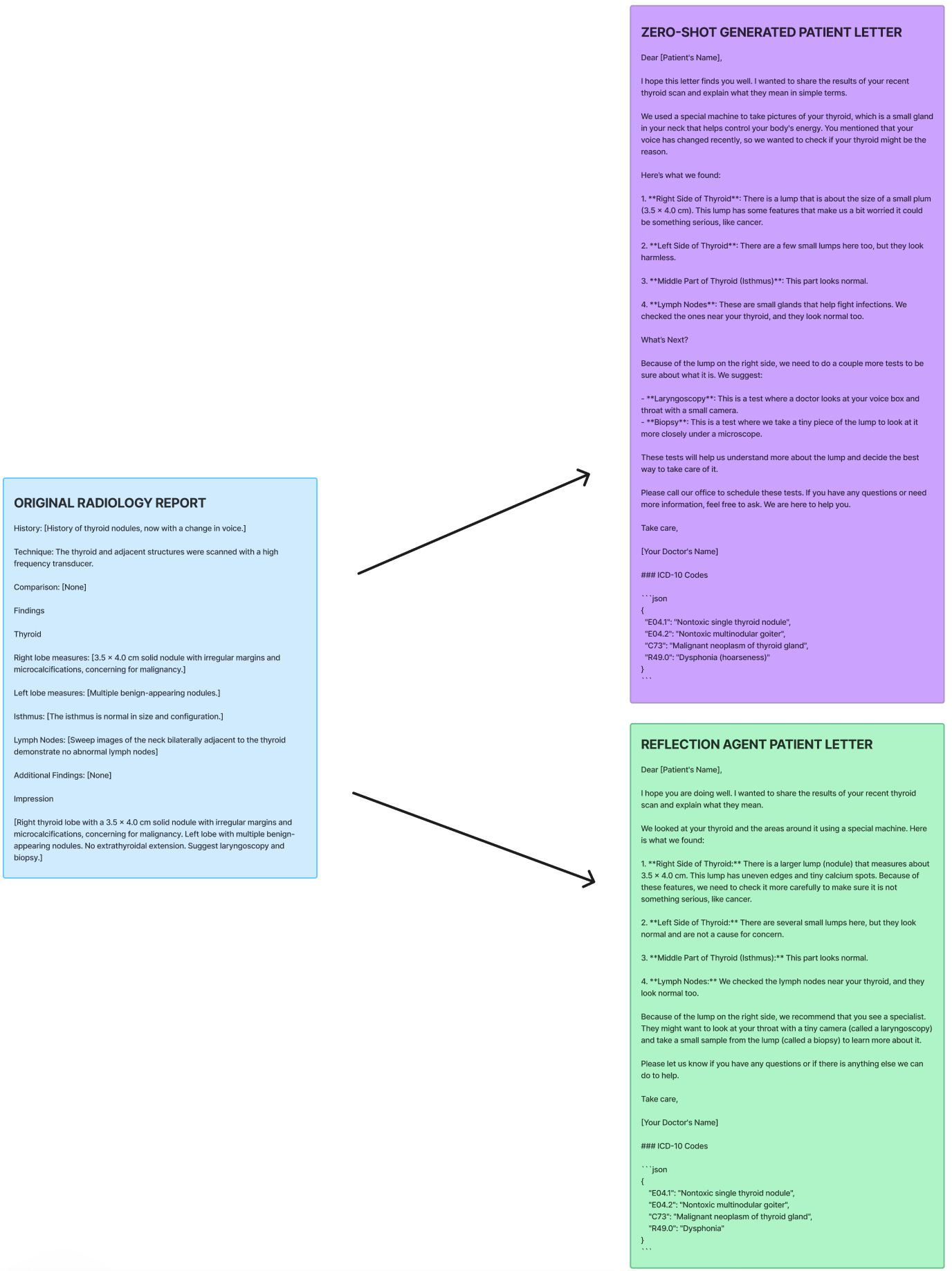}
  \caption{Patient-Friendly Letters generated from a US Thyroid Report}
  \label{fig:us_thyroid}
\end{figure}


\end{document}